\documentclass[preprint,amsmath,amssymb,superscriptaddress,longbibliography,aps,floatfix,footinbib]{revtex4-2}
\usepackage[]{graphicx}
\usepackage{tabularx}
\usepackage[usenames,dvipsnames]{color}
\usepackage{soul}
\usepackage{bm}
\usepackage{amsmath}
\usepackage{gensymb}
\usepackage[utf8]{inputenc}

\usepackage{amsfonts}
\usepackage{mathrsfs}
\usepackage{graphicx}
\usepackage{dcolumn}
\usepackage{bm}
\usepackage{color}

\usepackage[colorlinks,bookmarks=false,citecolor=blue,linkcolor=red,urlcolor=blue]{hyperref}
\usepackage{multirow}
\usepackage{physics}
\usepackage{siunitx}
\DeclareSIUnit{\barpressure}{bar}
\DeclareSIUnit\angstrom{\protect \text {Å}}
\usepackage{xcolor}

\begin{document}

\title{Pervasive electronic nematicity as the parent state of kagome superconductors}

\author{Muxian Xu}
\altaffiliation{These authors contributed equally to this work}
\affiliation{Department of Physics, Boston College, Chestnut Hill, Massachusetts 02467, USA}

\author{Siyu Cheng}
\altaffiliation{These authors contributed equally to this work}
\affiliation{Department of Physics, Boston College, Chestnut Hill, Massachusetts 02467, USA}

\author{Andrea Capa Salinas}
\affiliation{Materials Department, University of California Santa Barbara, Santa Barbara, California, USA}

\author{Ganesh Pokharel}
\affiliation{Perry College of Mathematics, Computing, and Sciences, University of West Georgia, Carrollton, GA 30118}
\affiliation{Materials Department, University of California Santa Barbara, Santa Barbara, California, USA}

\author{Alexander LaFleur}
\affiliation{Department of Physics, Boston College, Chestnut Hill, Massachusetts 02467, USA}

\author{Hong Li}
\affiliation{Department of Physics, Boston College, Chestnut Hill, Massachusetts 02467, USA}

\author{Hengxin Tan}
\affiliation{School of Physics and Astronomy, Shanghai Jiao Tong University, Shanghai 200240, China}

\author{Brenden R. Ortiz}
\affiliation{Materials Science and Technology Division, Oak Ridge National Laboratory, Oak Ridge, TN 37831, USA}

\author{Qinwen Deng}
\affiliation{Department of Physics, University of Pennsylvania, PA, USA}

\author{Binghai Yan}
\affiliation{Department of Physics, Pennsylvania State University: State College, Pennsylvania, USA}

\author{Ziqiang Wang}
\affiliation{Department of Physics, Boston College, Chestnut Hill, Massachusetts 02467, USA}

\author{Stephen D. Wilson}
\affiliation{Materials Department, University of California Santa Barbara, Santa Barbara, California, USA}

\author{Ilija Zeljkovic}\email{ilija.zeljkovic@bc.edu}
\affiliation{Department of Physics, Boston College, Chestnut Hill, Massachusetts 02467, USA}

\maketitle

\section{Abstract}
\textbf{Kagome superconductors $A$V$_3$Sb$_5$ ($A$ = Cs, K, Rb) have developed into an exciting playground for realizing and exploring exotic solid state phenomena. Abundant experimental evidence suggests that electronic structure breaks rotational symmetry of the lattice, but whether this may be a simple consequence of the symmetry of the underlying 2 $\times$ 2 charge density wave phase or an entirely different mechanism remains intensely debated. We use spectroscopic imaging scanning tunneling microscopy to explore the phase diagram of the prototypical kagome superconductor CsV$_3$Sb$_5$ as a function of doping. We intentionally suppress the charge density wave phase with chemical substitutions selectively introduced at two distinct lattice sites, and investigate the resulting system. We discover that rotational symmetry breaking of the electronic structure -- now present in short-range nanoscale regions -- persists in all samples, in a wide doping range long after all charge density waves have been suppressed. As such, our experiments uncover ubiquitous electronic nematicity across the $A$V$_3$Sb$_5$ phase diagram, unrelated to the 2 $\times$ 2 charge density wave. This further points towards electronic nematicity as the intrinsic nature of the parent state of kagome superconductors, under which other exotic low-temperature phenomena subsequently emerge.} 

\section{Introduction}
Kagome metals harbor rich phenomenology spanning different types of magnetism, charge ordering phases, electronic topology, and superconductivity in a variety of different compounds \cite{DiSante2025RMP}. Kagome superconductors $A$V$_3$Sb$_5$ have attracted particular attention on both theoretical \cite{Feng2021AVS, Tan2021kagome, Denner2021AVS, Lin2021AVS, Park2021kagome, Christensen2021AVS, Christensen2022AVS, Schwemmer2024kagome} and experimental \cite{Jiang2021KVS, zhao2021cascade, Qian2021CVS, Li2022STM_RSB, Guo2022chiralTransport, Wu2022CVS, Ptok2022AVS, chen2021roton, Liang2021_3DCDW, Li2021AVS, Kang2022CVS, Xiang2021two-foldCVS, yang_giant_2020, Xu2021STM, Uykur2022KVS, Zhou2021CVS, Broyles2022CVS, xu_three-state_2022, mielke2022time, Kato2022KVS, Wang2023strainSTM, Candelora2024kagome, Li2023pockets, Kautzsch2023CVS, Li2023STM_35K} fronts, driven in large part by the observation of superconductivity that appears to break time-reversal symmetry in the absence of magnetism and various states breaking different spatial symmetries \cite{b_ortiz_csv_3sb_5_2020, mielke2022time,Guo2022chiralTransport,Jiang2021KVS, PhysRevX.11.031050, Li2022STM_RSB, Li2023STM_35K, zhao2021cascade,xu_three-state_2022, Nie2022nematicity, Xiang2021two-foldCVS, asaba2024evidence, Li2022STM_RSB}. These phenomena are generally found to occur within the umbrella of the 2 $\times$ 2 charge density wave (CDW) phase, which forms in all stoichiometric members of the $A$V$_3$Sb$_5$ family below about 80-100 K \cite{b_ortiz_csv_3sb_5_2020, Li2021AVS}. Rotational symmetry breaking of the electronic structure is also detected around a similar temperature, in close proximity to the onset of the 2 $\times$ 2 CDW phase \cite{xu_three-state_2022, Li2023STM_35K, Xiang2021two-foldCVS} or slightly above \cite{asaba2024evidence}. As the 2 $\times$ 2 CDW phase breaks rotational symmetry by itself, largely driven by the shifts in the inter-layer phasing \cite{DiSante2025RMP}, a crucial question that remains is whether rotational symmetry breaking in the electronic structure is triggered by the symmetry of the 2 $\times$ 2 CDW, or whether its mechanism is entirely different.

In this work, we use chemical substitutions to intentionally suppress charge ordering phases in CsV$_3$Sb$_5$ and investigate the system using spectroscopic-imaging scanning tunneling microscopy (SI-STM). We visualize the evolution of electronic properties as a function of two different types of chemical substitutions -- Sn substituting at the Sb site and Ti substituting for V within the kagome plane. We discover that even after charge ordering states are fully suppressed, electronic state locally breaks six-fold rotational symmetry of the lattice, forming short-range ordered electronic nematic domains. These nanoscale domains show local anisotropy, which can be attributed to the anisotropic electron scattering and interference. Each domain appears to break all but one in-plane mirror symmetries. Our experiments demonstrate that electronic nematicity is present in a wide range of compositions across the phase diagram of CsV$_3$Sb$_5$, distinct from the symmetry of the 2 $\times$ 2 CDW, and as such reveal electronic nematicity as the unifying feature of kagome superconductor phase diagram.

\section{Results}

\begin{figure}
    \centering
    \includegraphics[width = \textwidth]{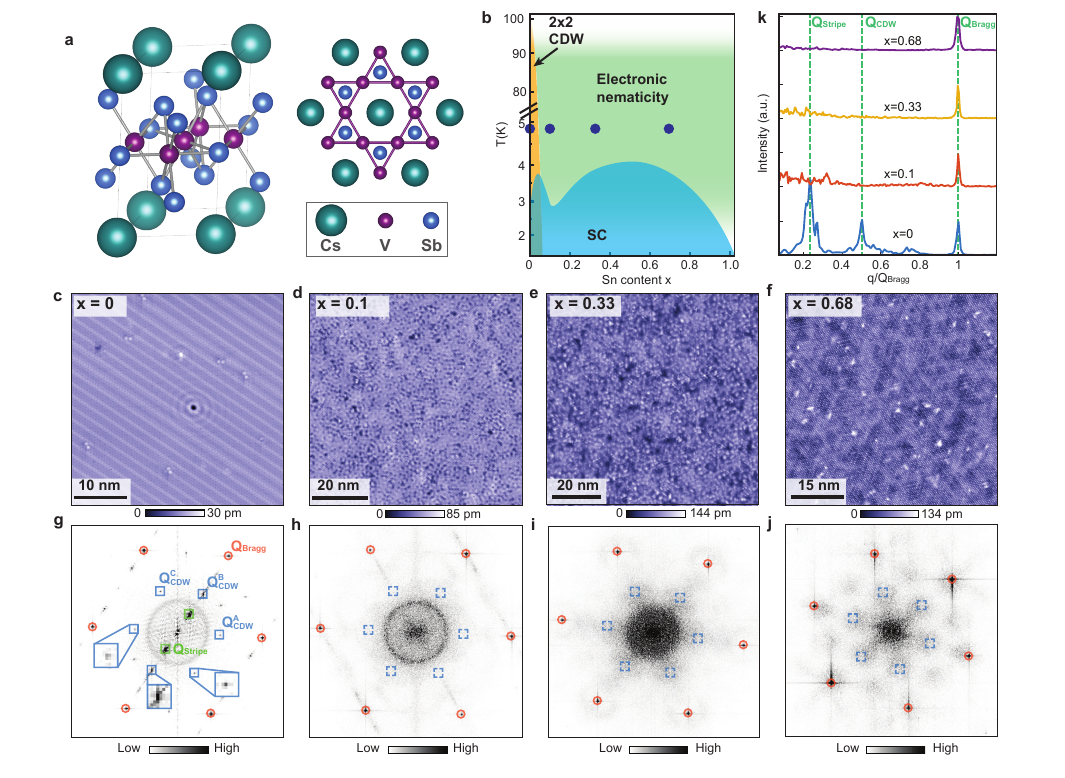}
    \renewcommand{\baselinestretch}{1}
    \caption{\textbf{Crystal structure and the evolution of the 2 $\times$ 2 CDW in CsV$_{3}$Sb$_{5-x}$Sn$_{x}$} - \textbf{a}, 3D ball model of the crystal structure of CsV$_{3}$Sb$_{5}$. \textbf{b}, A schematic phase diagram denoting the 2 $\times$ 2 CDW state and the superconducting state as a function of Sn doping. The four blue circles indicate the doping levels and temperatures at which the measurements in this study were performed. \textbf{c-f}, Representative STM topographs of the Sb-terminated surfaces for $x=0$ (\textbf{c}), $x=0.1$ (\textbf{d}), $x=0.33$ (\textbf{e}), $x=0.68$ (\textbf{f}). \textbf{g-j}, Fourier transforms (FTs) of the corresponding topographs in (\textbf{c-f}). FT peaks corresponding to the atomic lattice ($\textbf{Q}_{Bragg}$), the 2$\times$2 CDW ($\textbf{Q}_{CDW}$) and the 4$\times$1 charge stripes ($\textbf{Q}_{stripe}$) are labeled in \textbf{g}. In \textbf{h–j}, the atomic Bragg peaks and the expected CDW peak positions are marked using the same symbols as those in (g). 2 $\times$ 2 CDW peaks vanish for samples with $x \geq 0.1$. \textbf{k}, FT linecuts along the $\textbf{Q}_{stripe}$ direction (if present), showing the suppression of the 2 $\times$ 2 CDW peaks at higher Sn compositions. STM setup conditions: $V_{sample}$=100 mV, $I_{set}$=50 pA (\textbf{c}); $V_{sample}$=20 mV, $I_{set}$=200 pA (\textbf{d}); $V_{sample}$=20 mV, $I_{set}$=300 pA (\textbf{e}); $V_{sample}$=20 mV, $I_{set}$=200 pA (\textbf{f}).  } 
    \label{fig:1}
\end{figure}

We study bulk single crystals of CsV$_3$Sb$_5$, which exhibit a layered structure consisting of V$_3$Sb$_5$ blocks stacked between Cs layers (Fig.~\ref{fig:1}a). The crystals tend to cleave between the Cs layer and the Sb layer; we cleave the crystals at low temperature and in ultra-high vacuum, and promptly insert them into the STM head (Methods). Similarly to the majority of previous work on AV$_3$Sb$_5$ \cite{Jiang2021KVS, zhao2021cascade, chen2021roton, Liang2021_3DCDW, Li2022STM_RSB}, we focus on the Sb surface termination positioned directly above the kagome plane due to its structural stability and the direct access to bulk vanadium-derived kagome bands \cite{zhao2021cascade, chen2021roton}. We explore the evolution of electronic properties for different Sn compositions ($x=$0.15, $x=$0.33 and $x=$0.68) and Ti compositions ($y=$0.2) using spectroscopic imaging STM.

We first explore the Sn-doped CsV$_3$Sb$_5$ samples. From complementary bulk studies, the 2 $\times$ 2 CDW phase is rapidly suppressed at critical composition of $x_c \approx$ 0.06 (Fig.~\ref{fig:1}b) \cite{oey2022Sn-doping, liu2023Tidoping}. We confirm the suppression of the long-range 2 $\times$ 2 CDW in our STM topographs (Fig.~\ref{fig:1}c-f). Fourier transform (FT) of the STM topograph of the pristine compound shows the expected charge ordering states: the charge-stripe order around $\frac{1}{4} \textbf{Q}_{Bragg}$ along a single atomic Bragg peak direction and the 2 $\times$ 2 charge density wave phase around each $\frac{1}{2} \textbf{Q}_{Bragg}$ position (Fig.~\ref{fig:1}g). FT of the STM topograph of $x \sim 0.1$ sample  shows some signatures of unidirectional charge-stripe correlations as we reported in our previous work \cite{Kautzsch2023CVS}, but no 2 $\times$ 2 CDW peaks (Fig.~\ref{fig:1}d,h). The last two higher Sn compositions show distinctly different behavior -- sharp 2 $\times$ 2 CDW peaks are markedly absent, there is no charge-stripe ordering and the FTs look roughly symmetric (Fig.~\ref{fig:1}i,j). The absence of the 2 $\times$ 2 CDW can also be seen in the FT linecuts, starting from the FT center to the atomic Bragg peak for the four compositions, which show the absence of these peaks for representative STM topographs of all Sn samples studied here (Fig.~\ref{fig:1}k), thus confirming the expected behavior from bulk measurements \cite{oey2022Sn-doping, liu2023Tidoping}. Interestingly, we do observe very broad FT peaks around each $\frac{1}{2} \textbf{Q}_{Bragg}$ position for $x \ge 0.33$ (Fig.~\ref{fig:1}i,j), which are not there in the $x\approx0.1$ sample (Fig.~\ref{fig:1}h), and we explore this in more detail by spectroscopic imaging STM.

\begin{figure}
    \centering
    \includegraphics[width = \textwidth]{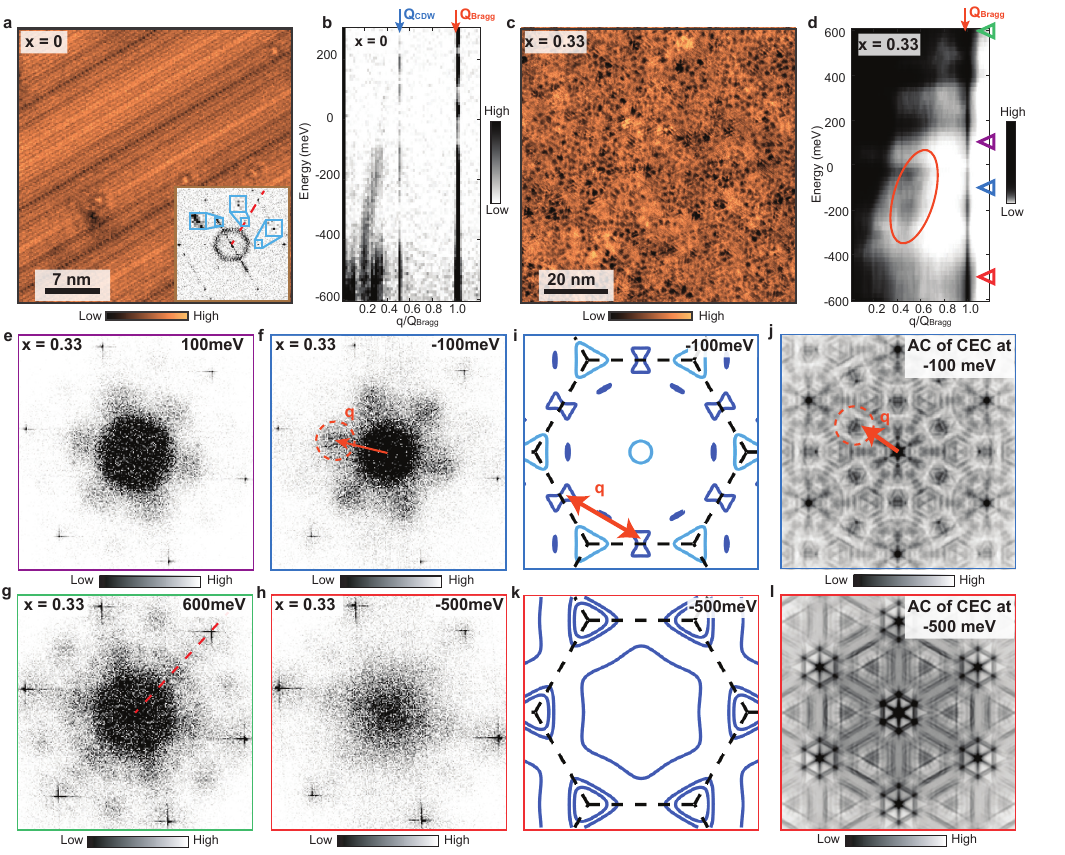}
    \renewcommand{\baselinestretch}{1}
    \caption{\textbf{Spectroscopic-imaging STM comparison of the $x = 0$ and $x = 0.33$ Sn-doped CsV$_3$Sb$_5$ samples.} - \textbf{a,c}, Representative differential conductance maps ($g(r,V) = dI(r,V)/dV$) for samples with $x = 0$ and $x = 0.33$. The inset in the bottom right of \textbf{a} shows the Fourier transform (FT) of a typical differential conductance map for $x = 0$. \textbf{b,d}, Waterfall plot of FT linecuts of the $g(r,V)$ maps for $x = 0$ and $x = 0.33$ samples, starting from the FT center towards one atomic Bragg peak. \textbf{e–h}, FTs of $g(r,V)$ maps at selected energies for the $x = 0.33$ sample. The corresponding energies are indicated by the colored triangles in \textbf{d}. The line cuts in \textbf{b} and \textbf{d} are taken along the red dashed lines shown in the insets of \textbf{a} and \textbf{g}. \textbf{i,k}, Constant energy contours (CECs) obtained from DFT calculations for $k_z=0$. \textbf{j,l}, Autocorrelation of the CECs in \textbf{i} and \textbf{k}, respectively. Orange arrow in \textbf{i} denotes the part in the CEC that leads to the scattering and interference pattern marked by the orange arrow in \textbf{f} and \textbf{j}. STM setup conditions: $V_{sample}$=20 mV, $I_{set}$=40 pA, $V_{exc}$=3 mV (\textbf{a}); $R_{tip-sample}$=1.5 G$\Omega$, $I_{set}$=200 pA, $V_{exc}$=4 mV (\textbf{b}); $V_{sample}$=20 mV, $I_{set}$=300 pA, $V_{exc}$=1 mV (\textbf{c}); $R_{tip-sample}$=0.6 G$\Omega$, $I_{set}$=1 nA, $V_{exc}$=20 mV (\textbf{d, e-h}).  }
    \label{fig:2}
\end{figure}

To establish a baseline, we note that FTs of d$I$/d$V$ maps of the pristine sample show a set of six crisp peaks at $\frac{1}{2} \textbf{Q}_{Bragg}$ (approximately 1-2 pixel width) across a wide energy range, corresponding to the 2 $\times$ 2 CDW phase (Fig.~\ref{fig:2}a,b). Although the equivalent sharp FT features are not present in the $x = 0.33$ sample (Fig.~\ref{fig:2}d), we do however observe the emergence of broad FT peaks (spanning $\sim$20$\%$ of $\textbf{Q}_{Bragg}$) near the $\frac{1}{2} \textbf{Q}_{Bragg}$ position at some energies (Fig.~\ref{fig:2}d,e,f). We rule out residual doping-induced short-range 2 $\times$ 2 CDW as the origin of these modulations for the following reasons. First, no $\frac{1}{2} \textbf{Q}_{Bragg}$ FT peaks are detected in the FTs of topographs or d$I$/d$V$ maps at the intermediate doping of $x\sim0.1$, just as the long-range 2 $\times$ 2 CDW is suppressed (Extended Data Fig.~1). Second, $\frac{1}{2} \textbf{Q}_{Bragg}$ peaks are only observed at some energies, unlike those in the pristine sample observed ubiquitously across a wide energy range (Fig.~\ref{fig:2}b). Third, the modulation wave vector shows slight dispersion and notably changes in shape as a function of energy (Fig.~\ref{fig:2}d-h, Extended Data Fig.~2). Based on this, we attribute these features to the electron scattering and interference, similarly to what has been studied in more detail in the pristine compound \cite{zhao2021cascade, Li2023STM_35K, Li2023pockets}. 

To substantiate this further, we turn to density functional theory (DFT) calculations (Methods). Constant energy contours of the $x\sim0.25$ system show a number of evolving features, which at some energies result in small pockets around $M$ points (Fig.~\ref{fig:2}i). These pockets are nicely connected by exactly $\frac{1}{2} \textbf{Q}_{Bragg}$ wave vector, which would give rise to the scattering wave vectors seen in FTs of d$I$/d$V$ maps (wave vector $q$ in Fig.~\ref{fig:2}f,i,j). We perform mathematical auto-correlation of the constant energy contours to approximate the FT of STM d$I$/d$V$ maps, assuming all scattering processes are possible, which shows the $\frac{1}{2} \textbf{Q}_{Bragg}$ peaks (Fig.~\ref{fig:2}j). As the band structure evolves with energy, the pockets around $M$ evolve as well, so at other energies we no longer observe the same features around $\frac{1}{2} \textbf{Q}_{Bragg}$ (Fig.~\ref{fig:2}h,k,l).

We proceed to examine the morphology of the real-space electronic signal in more detail. Interestingly, we find that the d$I$/d$V$ map where $\frac{1}{2} \textbf{Q}_{Bragg}$ wave vectors are strong is composed of puddles, each showing locally anisotropic charge modulations most pronounced along one lattice direction (Fig.~\ref{fig:3}a). For instance, within each of the three highlighted regions, we can see that the modulations are seemingly oriented along a single crystalline directions (Fig.~\ref{fig:3}b). The wave vector associated with the modulations is about $\frac{1}{2} \textbf{Q}_{Bragg}$ as it can be seen in the corresponding FTs of the zoomed-in topographs. So while the FT of the entire field-of-view may appear roughly six-fold symmetric (i.e. Fig.~\ref{fig:2}e), it is composed of short-range regions that each locally break rotational six-fold symmetry of the lattice. Each individual region breaks two in-plane mirror symmetries but appears to generally preserve the in-plane mirror symmetry along the direction of the arrows. This is consistent with the highest Sn composition studied at $x=$0.68, where we can again observe domains with local electronic anisotropy (Fig.~\ref{fig:3}c,d).

To investigate whether the rotational symmetry breaking measured is dependent on the particular chemical substitution used, or whether this is a general feature of the kagome superconductor phase diagram, we study the effects of another type of dopant, Ti substituting for V in CsV$_{3-x}$Ti$_x$Sb$_5$ bulk single crystals (Fig.~\ref{fig:4}). Based on bulk scattering experiments, 2 $\times$ 2 CDW is expected to be suppressed at Ti composition higher than $y_c \sim 0.05$ \cite{liu2023Tidoping}. Consistent with this, in the $y=0.2$ Ti-doped sample studied here, 2 $\times$ 2 CDW peaks are absent in the FTs of STM topographs (Fig.~\ref{fig:4}a,b). Although we have not been able to perform momentum-resolved imaging using electron scattering and interference as that in the Sn substituted samples, STM topographs of the Ti-doped samples show another intriguing feature. Namely, we observe local anisotropy of the atomic lattice modulations, characterized by modulations being substantially stronger along one lattice direction. Fig.~\ref{fig:4}a shows one region of the sample where anisotropy of the electronic modulations rotates from $\textbf{Q}_{Bragg}^B$ to $\textbf{Q}_{Bragg}^C$ direction across the domain boundary (dashed orange line in Fig.~\ref{fig:4}a). Zoom-ins on each domain and associated Fourier transforms depict this more clearly (Fig.~\ref{fig:4}c,d). We also show the amplitude maps associated with corresponding atomic Bragg peak over the same area (Figs.~\ref{fig:4}e,f). For instance in Fig.~\ref{fig:4}e-f, it can be seen that the lower right area lights up in for the $\textbf{Q}_{Bragg}^B$ amplitude map, while the upper left area has high intensity for the $\textbf{Q}_{Bragg}^C$ direction. Extended Data Fig.~3 shows another nematic boundary on the same sample, this time between $\textbf{Q}_{Bragg}^A$ and $\textbf{Q}_{Bragg}^C$ directions. From the symmetry perspective, our observations on the Ti-doped sample are consistent with those on the Sn-doped samples -- we observe nanoscale electronic nematic domains, which locally break the rotational symmetry of the lattice. 

\section{Discussion}
Extensive experimental body of work on $A$V$_3$Sb$_5$ has detected rotational symmetry breaking of the electronic structure around the CDW onset temperature \cite{xu_three-state_2022, Li2023STM_35K, Xiang2021two-foldCVS} or slightly above the CDW transition \cite{asaba2024evidence}, and as such it has remained unclear to what extent rotational symmetry breaking may be a simple consequence of the underlying symmetry of the 2 $\times$ 2 CDW phase. To shed light on this issue, we studied CsV$_3$Sb$_5$ in a wide range of compositions to systematically suppress charge ordering states. Our experiments reveal short-range $q=0$ electronic nematicity across the phase diagram of CsV$_3$Sb$_5$, long after the 2 $\times$ 2 CDW and other charge ordering states are suppressed at substantially lower doping. The phenomenology is confirmed in two different types of samples, chemically doped with either Sn or Ti, thus suggesting this could be a generic feature of the kagome superconductor phase diagram. The electronic nematic state is short-range, with local anisotropic modulations oriented along different crystalline directions forming three different types of domains. Within each nematic domain, modulations are $C_2$-symmetric, with two in-plane mirror lines that are clearly broken, and one that appears to be intact. The short-range nature of electronic nematicity in doped samples likely arises due to increased disorder caused by the dopants, but the relationship between the two should be examined more closely in future work. 

\begin{figure}
    \centering
    \includegraphics[width = \textwidth]{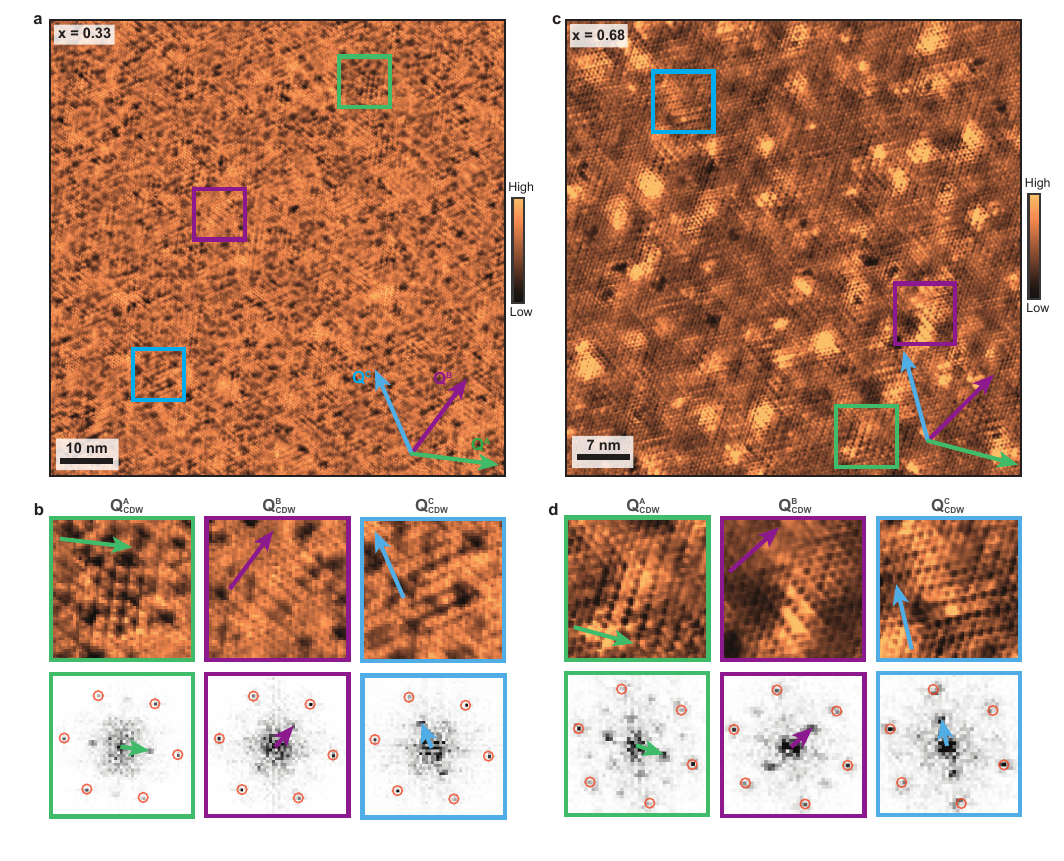}
    \renewcommand{\baselinestretch}{1}
    \caption{\textbf{Nanoscale domains with anisotropic modulations.} \textbf{a,c}, Representative high-resolution d$I$/d$V$(\textbf{r}, $V$) maps with $x = 0.33$ and $x = 0.68$. \textbf{b,d}, Selected regions of the d$I$/d$V$(\textbf{r}, $V$) maps (top row) and their Fourier transforms (FTs) (bottom row). Each selected region is dominated by the charge modulation along one crystalline direction (marked by the colored arrows) being substantially stronger compared to other directions. Scattering peaks marked by the colored arrows in the FTs further demonstrate the nematic nature of charge modulation with about $\frac{1}{2}\textbf{Q}_{Bragg}$ wave vector. STM setup conditions: $V_{sample}$=-200 mV, $I_{set}$=500 pA, $V_{exc}$=5 mV (\textbf{a}); $V_{sample}$=50 mV, $I_{set}$=200 pA, $V_{exc}$=10 mV (\textbf{c}).  }
    \label{fig:3}
\end{figure}

\begin{figure}
    \renewcommand{\thefigure}{4}  
    \renewcommand{\figurename}{FIG.}  
    \centering
    \includegraphics[width = \textwidth]{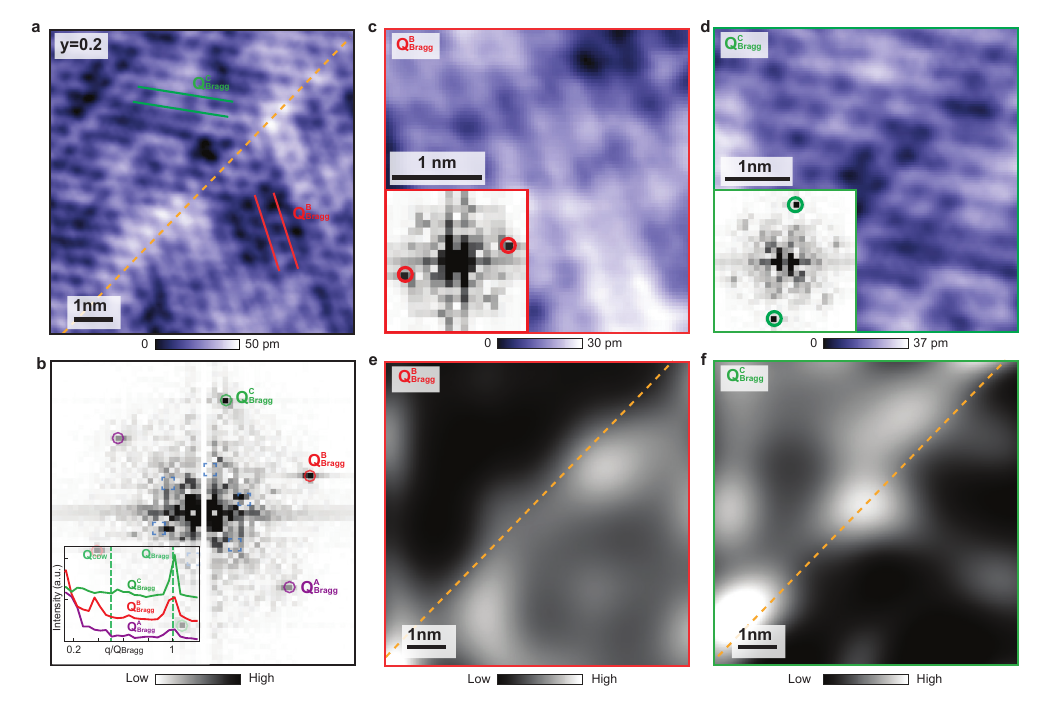}
    \renewcommand{\baselinestretch}{1}
    \caption{\textbf{Electronic nematic domains in the Ti-doped CsV$_{3-y}$Sb$_5$Ti$_y$ samples.} - \textbf{a}, Representative STM topograph of CsV$_{3-y}$Sb$_{5}$Ti$_{y}$ with $y=0.2$, and \textbf{b}, Fourier transform (FT) of the area in (a). Colored circles mark the atomic Bragg peaks, while the blue squares mark the expected 2 $\times$ 2 CDW peak locations. The inset in the bottom left of \textbf{b} shows the FT line cuts of \textbf{b}, starting from the FT center towards each atomic Bragg peaks, showing the absence of 2 $\times$ 2 CDW. \textbf{c,d}, STM topographs of individual nematic domains. The insets in the bottom left of \textbf{c} and \textbf{d} are their FTs, with the atomic Bragg peaks with the strongest intensity marked by the colored circles. \textbf{e,f}, Amplitude map of the lattice modulations obtained by applying a Fourier-space filter that isolates a single pair of atomic Bragg peaks. Enhanced intensity in this map corresponds to regions where the lattice modulation is locally stronger along this direction. STM setup conditions: $V_{sample}$=100 mV, $I_{set}$=100 pA (\textbf{a,c,d}).  }
\label{fig:4}
\end{figure}

The observation of domains oriented in different directions also provides crucial evidence that electronic anisotropy observed is not STM tip induced, since images (such as those in Figs.~\ref{fig:3} or Fig.~\ref{fig:4}) are scanned using the same tip. We also note that the FT of a larger field of view inevitably depends on how many domains of each type it encompasses. For instance, the FT may look roughly six-fold symmetric if all three types of domains are equally prevalent, or at the other extreme the FT may appear to break all mirror symmetries if areas encompassing the three types are all inequivalent. This highlights the necessity for atomic-scale examination of the samples to pin down the underlying symmetries. 

Pristine CsV$_3$Sb$_5$ exhibits abundant intriguing phenomena at low-temperature -- charge orders, unusual superconductivity and time-reversal symmetry breaking -- all of which may arise from the electronic nematic state imaged here that clearly persists in their absence in the vast portion of the phase diagram. It would be interesting to explore if and how this phase evolves into a chiral state at low temperature by breaking the remaining in-plane mirror symmetry, provided that tip anisotropy \cite{Candelora2025} and strain \cite{Wang2023strainSTM} as extrinsic factors can be excluded. Lastly, we note that long-range electronic nematicity has been observed in cousin kagome superconductor CsTi$_3$Bi$_5$ \cite{Li2023CTB_nematicity, Yang2024CTB135_STM}, with the same local symmetry as that reported here, thus painting a unified picture of electronic nematicity as the parent state of kagome superconductors.\\

\textit{Note: During the preparation of the manuscript, we became aware of related work in Ref.~\cite{Huang2025chiral-nematic} on Ti-doped CsV$_3$Sb$_5$.}\\

\noindent{\bf Methods}\\
\indent \textit{\textbf{Sample growth:}} 
Single crystals of CsV$_3$Sb$_{5-x}$Sn$_x$, x$\sim$0.33 and x$\sim$0.68  were synthesized by the self-flux growth method. Elemental Cs (Alfa 99.98 $\%$), V powder (Sigma 99.9 $\%$), Sb shot (Alfa 99.999$\%$), and Sn shot (99.999 $\%$) were weighed out in $25 : 18.75 : 80 : 40$ and $25:18.75:60:60$ ratios respectively under an argon environment -- H$_2$O and O$_2$ $<$ 0.5 ppm -- and milled in a SPEX8000D mixer inside a tungsten carbide vial for 60 min. The milled powder was packed in alumina crucibles and sealed inside a steel tube for growth. The growth sequence consisted of a heating step up to 1000 $\degree$C, dwelling at this temperature for 12 hours, cooling down to 950 $\degree$C at 5 $\degree$C/hour and further slow cooling at 1 $\degree$C/hour down to 500 $\degree$C. The resulting shiny hexagonal crystals were manually extracted at room temperature in air.

Single crystals of CsV$_{3-x}$Ti$_x$Sb$_5$ (x=0.2) were grown via a conventional flux-based growth technique. Vanadium powder (Sigma-Aldrich, purity level 99.9$\%$) was further cleaned using a mixture of Hydrochloric acid and isopropyl alcohol to remove residual oxides. Cs (liquid, Alfa 99.98$\%$), cleaned V, Ti (powder, Alfa 99.9$\%$), and Sb (shot, Alfa 99.999$\%$) were loaded inside a tungsten carbide milling vial with a molar ratio of Cs$_{20}$V$_{12}$Ti$_{3.5}$Sb$_{120}$ and then milled for about an hour. The milled powder was sealed inside a stainless-steel tube in an argon atmosphere. The samples were heated at a rate of 200 $\degree$C/h to 1000 $\degree$C and held there for 10 h. Then the tubes were cooled to 900 $\degree$C at 25 $\degree$C/hr and then to 500 $\degree$C at 1 $\degree$C/hr. Plate-like, shiny single crystals were manually separated from the flux after the tube had naturally cooled to room temperature from 500 $\degree$C.

\textit{\textbf{STM experiments:}}
Bulk single crystals CsV$_3$Sb$_{5-x}$Sn$_x$ and CsV$_{3-x}$Ti$_x$Sb$_5$ were cleaved at cryogenic temperature and in ultra-high vacuum, and immediately inserted into the STM head. STM data were acquired using a customized Unisoku USM1300 microscope at the base temperature of about 4.5 K. Spectroscopic measurements were made by using a standard lock-in technique with a 910 Hz frequency and bias excitation as detailed in the figure captions. STM tips used were home-made chemically-etched tungsten tips.

\textit{\textbf{DFT calculations:}} 
DFT calculations were performed using the Vienna \textit{ab-initio} Simulation Package (VASP) \cite{Kresse1996DFT}. The exchange--correlation interaction was treated within the generalized gradient approximation using the PBE parameterization \cite{Perdew1996DFT}. A plane-wave energy cutoff of 300~eV was employed, and all self-consistent calculations were carried out using a $12 \times 12 \times 6$ \textit{k}-point mesh. Since previous studies have reported that even a small amount of Sn doping suppresses the CDW, all first-principles calculations were performed on the pristine phase. Spin--orbit effect was included in all calculations.

To obtain the electronic structure of Sn-doped CsV$_3$Sb$_5$, we first constructed tight-binding Hamiltonians for pristine CsV$_3$Sb$_5$ and CsV$_3$Sb$_4$Sn, where our total-energy calculations indicate that Sn preferentially occupies the Sb site within the kagome plane. The maximally localized Wannier function method \cite{Pizzi2020DFT} with the V $d$ and Sn/Sb $p$ orbital as the projection basis was employed. The Hamiltonian of CsV$_3$Sb$_{5-x}$Sn$_x$ was then generated in real space by linearly interpolating hopping parameters of the same orbitals and atomic sites in the two Hamiltonians,
\[
t_x = (1-x)\,t_0 + x\,t_1,
\]
where $t_x$, $t_0$, and $t_1$ denote the hopping amplitudes for CsV$_3$Sb$_{5-x}$Sn$_x$, CsV$_3$Sb$_5$, and CsV$_3$Sb$_4$Sn, respectively. The Fermi surface of CsV$_3$Sb$_{5-x}$Sn$_x$ was subsequently extracted from the resulting band structure on a dense $150 \times 150 \times 80$ \textit{k}-point grid in the three-dimensional Brillouin zone.

\noindent{\bf Data availability}\\
The data that support the findings of this study are available from the corresponding authors upon reasonable request.\\
\\
\noindent{\bf Code availability}\\
The code that supports the findings of the study is available from the corresponding authors upon reasonable request.\\
\\
\noindent{\bf Correspondence and requests for materials} should be addressed to
\textit{ilija.zeljkovic@bc.edu}.\\
\\
\noindent{\bf Competing financial interests}\\
The authors declare no competing financial interests.\\

\bibliographystyle{custom-style.bst}
\bibliography{biblio_1_2_cleaned_subset_check_by_hand.bib}

\newpage

\end{document}